\documentstyle[multicol,aps,psfig,epsf]{revtex}

\begin{document}
\title{Internal avalanches in models of granular media}
\author{Supriya Krishnamurthy$^a$, Hans Herrmann$^{a,b}$
Vittorio Loreto$^a$, Mario Nicodemi$^c$ and Stephane Roux $^d$}
\pagestyle{myheadings}
\address{
$a)$ P.M.M.H. Ecole Sup\'erieure de Physique et Chimie Industrielles, \\
10, rue Vauquelin, 75231 Paris CEDEX 05 France \\
$b)$ ICA1, Univ. Stuttgart, Germany \\
$c)$ Dipartimento di Fisica, Universit\'a di Napoli ``Federico II'',
Unit\`a INFM and INFN Napoli \\
Mostra d'Oltremare, Pad. 19, 80125 Napoli, Italy, \\
$d)$ Laboratoire Surface du Verre et Interfaces, Unit\'e Mixte de Recherche
CNRS/St-Gobain, \\
 39, Quai Lucien Lefranc, F-93303 Aubervilliers Cedex, France.}

\maketitle
\date{\today}
\begin{abstract}
We study the phenomenon of internal avalanching within the context
of recently introduced lattice models of granular media. The avalanche
is produced by pulling out a grain at the base of the packing and
studying how many grains have to rearrange before the
packing is once more stable. We find that the avalanches are
long-ranged, decaying as a power-law. We study the distriution of
avalanches as a function of the density of the packing and find that
the avalanche distribution is a very sensitive structural probe
of the system.
\end{abstract}
\smallskip
\vskip2pc

\section{Introduction}

The internal structure and geometry of granular 
packings are very different from those of other systems such as
liquids or solids and a lot of work has been devoted to understand
them \cite{grain}. In particular, surface avalanches
occurring in granular packings have been extensively studied
\cite{grain}. 
In this paper we look at another sort of avalanching
phenomenon also indicative of the internal structure - the phenomenon
of internal avalanching occurring under small perturbations.

Recently simple lattice gas models have been proposed to describe slow 
dynamical processes in granular media, models whose basic ingredient
is the geometric frustration in particle motion 
\cite{conihermann,nicohans1,tetris}. 
These models reproduce experimentally observed phenomena such as 
slow relaxation in compaction, segregation\cite{segtet}, 
experimental irreversible-reversible cycles as well as 
the presence of ``aging'' and glassy behavior\cite{nicodemi}. 

Within the context of these lattice gas models, we study 
in the present paper the features of internal avalanching. 
In particular we observe how the internal structure of the 
packing is reorganized after 
a small  perturbation, such as pulling out a grain from the base. 
We find that surprisingly, the packing can undergo large
rearrangements even under such a minimal perturbation.  
The size distribution of the 
produced events is very broad, being a power law over several 
orders of magnitude. This shows the strong sensitivity of 
packings, in the above models, to small perturbations which may 
trigger huge events up to the scale of the system itself. We study the
avalanche distribution as a function of the density of the packing and
find that the distribution shows a strong dependence on density,
indicating that it is a very sensitive structural
probe of the system.

The outline of the paper is as follows. In the following section, we
define the model which we have studied numerically.
In section $III$ we make precise
the definition of an internal avalanche and define the 
quantity studied.
Our results are described in section $IV$ and
the conclusions and discussion follow in section $V$.

\section{{\em Tetris}-like models}

In this section we briefly review the definitions and some basic 
properties of the {\em Tetris}-like lattice models \cite{tetris} 
used in our simulations. 
The choice of the name ``Tetris''
captures the original idea 
of the computer game, of difficult parking problems in a packing of 
objects of different shapes. Frustration arises in granular packings 
owing to the interlocking of grains having different shapes.  Different
shaped particles have different sorts of excluded volume effects which
leads to frustration in the packing.
This
geometrical feature is captured in this class of models. 
Hence, in this model,
the complexity of the problem lies in the complexity of 
the particle arrangements in the packing. 
In a very general way we can define the
model with a complex interaction matrix which tells us, for each particle, 
what are the constraints on the combinations of 
particles around it.  For the sake of simplicity, in what follows, 
we define and study the model in its simplest version 
with two kinds of elongated particles.

In order to describe various experimentally observed properties of
granular media,
another class of frustrated lattice gas models with quenched disorder 
has also been introduced \cite{conihermann,nicohans1}. Although in
this paper we just focus on the {\em Tetris}-like models, it would be
interesting to  check the same phenomenology in this other class
of frustrated models as well. 

\begin{figure}[h]
\centerline{
        \psfig{figure=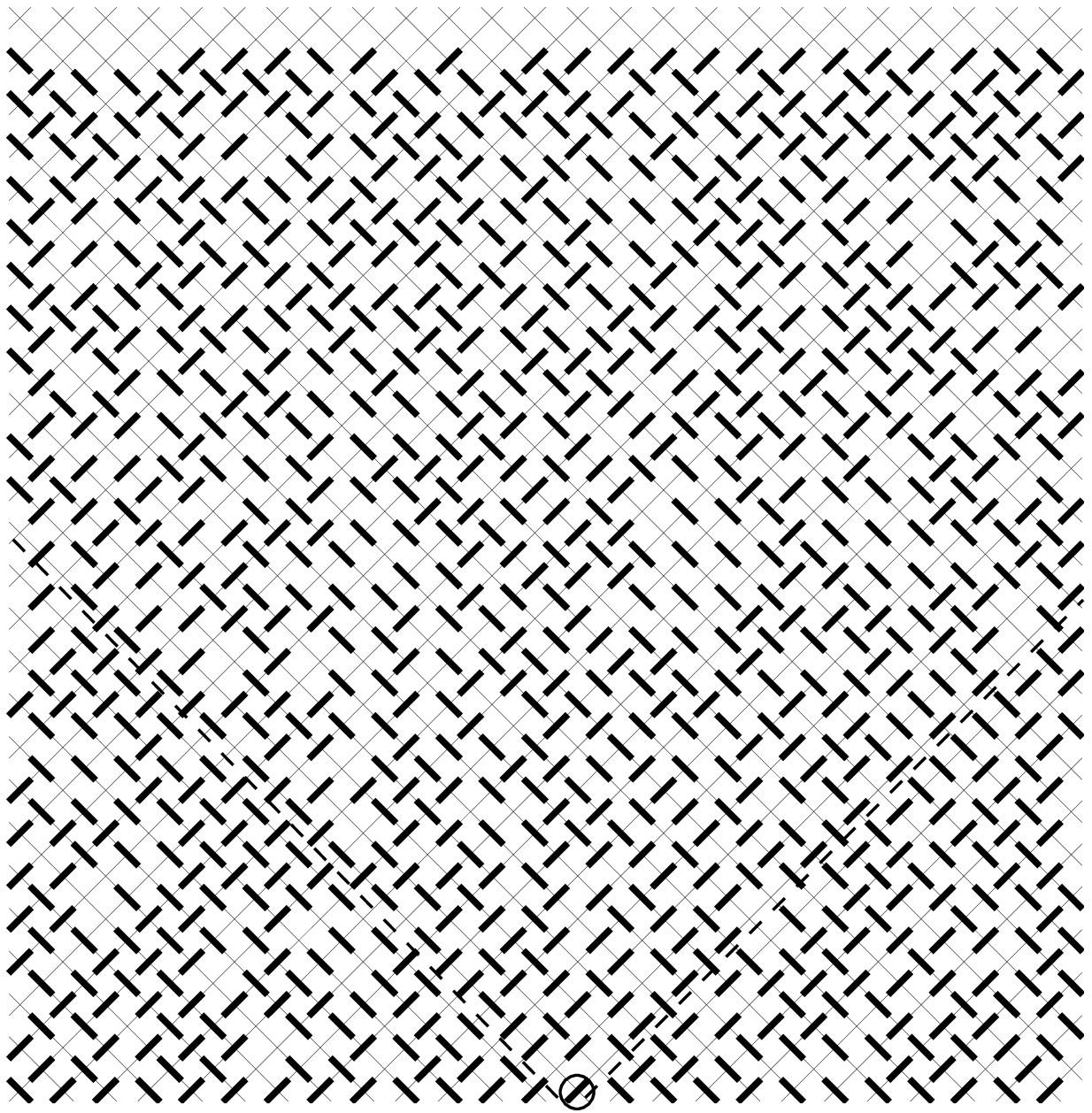,width=12cm,angle=0}}

        \vspace*{0.5cm}
\caption{}
\label{figmodel}
\end{figure}

The simplest version of the {\em Tetris}-like model can be defined 
by considering a system of particles which occupy the sites of a 
square lattice tilted by $45^0$ (see Fig.~\ref{figmodel}),  
with periodic boundary conditions in the horizontal direction
(cylindrical geometry) and a rigid wall at the bottom.
Particles cannot overlap and this condition produces
very strong constraints (frustration) on their relative positions. For
instance in the simplest case of two kinds of elongated
particles pointing in two (orthogonal) directions, the frustration
implies that two identical particles (pointing in the same direction)
cannot occupy neighboring sites in this direction (Fig. ~\ref{figmodel}).
There is no other form of 
interaction between particles, and in this sense the model is 
purely geometrical.
The system is initialized by filling the container, by 
inserting the grains at the top of the system,
one at a time, and letting them fall under gravity.
The particles perform
an oriented random walk on the lattice until they reach 
a stable position defined as a position in which they cannot fall
any further because of other particles below them.

The density reached by this filling procedure is $\sim 0.747$ and is
the lowest density a random packing can reach in this model with
two kinds of elongated particles
(the density of a packing is measured by averaging over the densities
of each row in the lower half of the system).
Higher
densities are reached by ``shaking'' the system, a procedure described
below. This procedure has been studied earlier in relation to the
experimentally observed slow density increase which occurs in these
models \cite{tetris} as the shaking continues.
In this paper however we concern ourselves with
probing the changes in the
structure of the packing caused by the shaking procedure,
by studying internal
avalanching as a function of the density. 

The effect of vibrations is introduced by the possibility of  
the particles  moving up with a probability 
$p_{up}$ and of moving down with a 
probability  $p_{down}=1-p_{up}$  
The quantity $1/\ln \frac{1}{x}$, 
with $x=p_{up}/p_{down}$, plays  the role of an effective temperature 
and can be related to the tap intensity amplitude.
The shaking procedure we use can be divided into two alternating steps.  
First, in a {\em heating} process (tapping)
the system is perturbed by putting $p_{up}\ne 0$ and performing 
a fixed number $N$ of attempts of movements per particle.
At the end of this step, due to a nonzero $p_{up}$,
the system is in an unstable state with
many particles in positions unsupported by particles below.
We now allow the system to relax under gravity
setting $p_{up}=0$.
The relaxation process ({\em Cooling}) is terminated
when all the particles once more acquire stable positions.
The system is now in a static state and the process of heating
and relaxing the system is repeated a specified number of times. 
The density the system reaches depends, on average, on this number. 
The basic features of our model are robust with
respect to variations in the exact  Monte-Carlo procedure used.

This simple version of the Tetris model presents a trivial 
``antiferromagnetic'' ground state. That is, the highest density
packing that the system can reach with the above tapping procedure 
always has $ \rho =1$ and is ordered in one of two possible
orderings : Even(odd) rows consist of rods with $+45^{0}$($-45^{0}$) 
orientations. A state with only one kind of ordering is hence called a
``single-domain'' state.
The existing of only two distinct orderings
is potentially a drawback since 
a real granular system  contains much more disorder 
due to a wider shape distribution and the absence of a lattice. 
In order to incorporate this effect, 
the Tetris model can be modified by considering particles
with more complicated shapes\cite{tetris,segtet}. 
This prevents the occurrence of an ordered
ground state. In this paper however we study exclusively the simpler
model described, with rods of two different orientations, with a brief
discussion of the more complicated case in section $V$.

\section{Internal avalanches}

We study the effects of small perturbations on packings as a function
of density, by studying in detail the phenomenon of internal
avalanching 
within the scope of the models described above.
Specifically we focus our attention on the rearrangements of grains
generated in a static assembly by the extraction of a grain at the
base.  
The creation of a void in the lattice destabilizes
the neighboring grains above it. One of these may then fall down to
fill the void,  
if the geometry of the packing allows the motion (i.e., if the 
orientation of the grain fits the local conformation). In this
case, the  
net effect is that the void propagates one lattice step upwards 
destabilizing its neighbors in the layer above and so on. 
How effective this process is in causing the restructuring of the 
configuration depends on the precise structure of the packing.
That large scale restructuring events are indeed
possible is reflected in the fact that there are certain local
configurations in which the motion of a single particle results in the
motion of two particles above it, thus creating a second moving void. 
As a result of this ``birth'' process, moving voids can not only
propagate up or get trapped but also  multiply and hence 
lead to large avalanches.

\begin{figure}[h]
\centerline{
        \psfig{figure=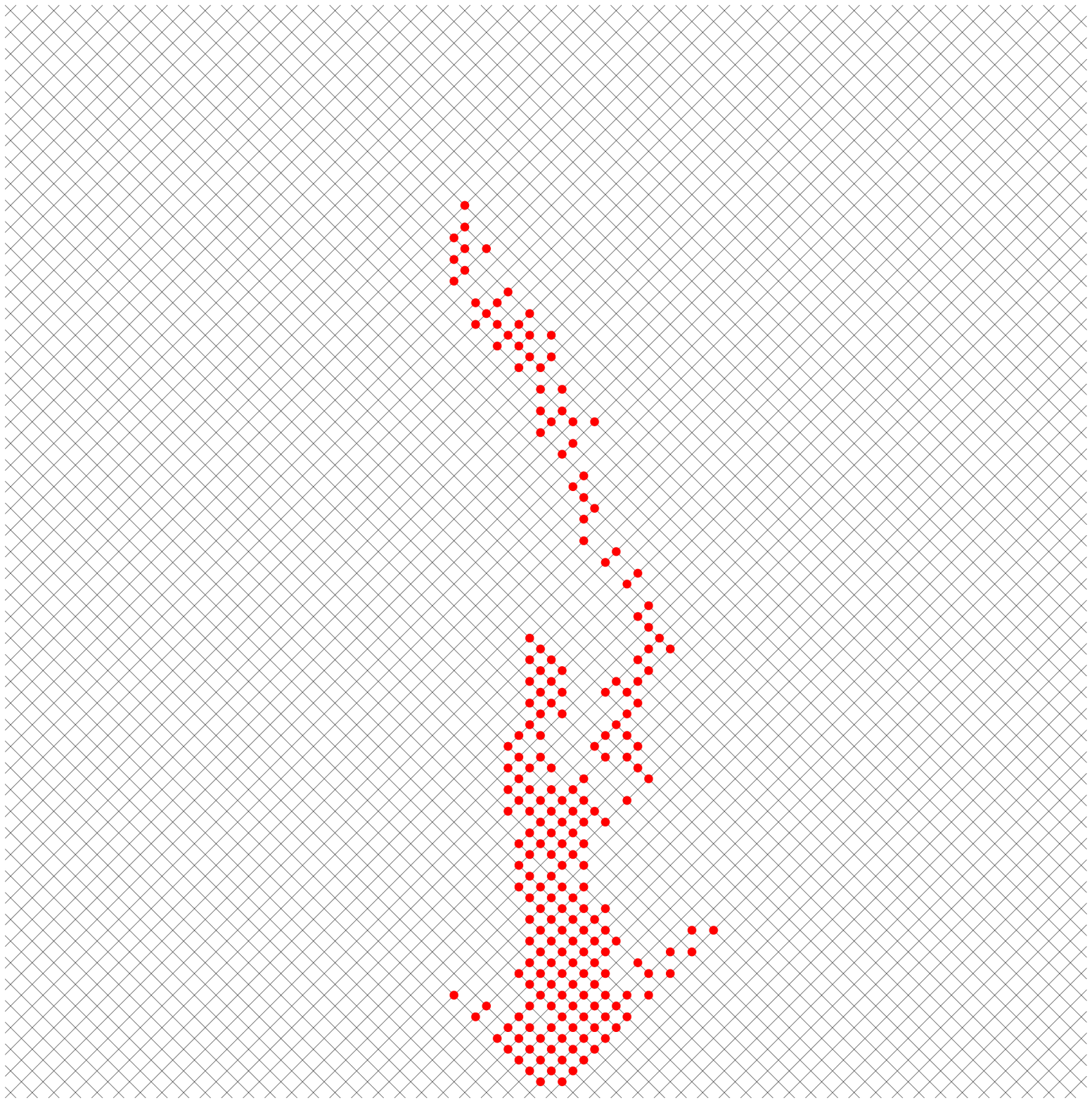,width=12cm,angle=0}}

        \vspace*{0.5cm}
\caption{}
\label{figaval}
\end{figure}

We begin by preparing the system in the loose packed state by the
procedure described in the previous section.
As mentioned earlier, in this way the system attains  
in average a density of about $\rho=0.747$ ($\rho_{ld}$).
The avalanche distribution
in this state is studied by the following 
procedure. An initial state at the loose packed density is produced 
and a particle is randomly removed from the base. The total number of
grains that move as a result of this removal is then calculated by
checking the system row by row and letting all unstable particles
settle under gravity. We have also checked that choosing unstable
particles completely stochastically rather than row by row does not
change the results quantitatively.
By invoking the rules
of stability introduced in this model, it is easy to see that when a
particle is removed from the packing, it at most destabilizes
all the particles within a cone with its apex at the removed
particle (Fig. ~\ref{figmodel}). Statistics for the avalanches is
obtained by repeating this process of counting the number
of unstable grains for various other initial states at the 
loose packed density.
Thus the avalanche distribution obtained from
this ensemble averaged procedure is indicative of this particular
density. To obtain the distribution at  higher densities the same
ensemble averaging procedure is followed where now each member of the
ensemble is generated by starting from an initial state at the loose
packed density followed by a specified number of shakes. For each
member of the ensemble, the magnitude of the internal avalanche is
studied by counting as before the total number of  particles
rearranged as a result of removing one particle from the last row. The
density that parametrises this avalanche distribution is 
just the average density of the ensemble.
In everything that follows unless otherwise mentioned,
this is the procedure we used to generate
avalanche distributions at different densities, by averaging over
statistics obtained for $100,000$ different initial conditions. 
A figure of an actual
avalanche is shown in Fig ~\ref{figaval}. The shaded dots represent
the original positions of particles which moved as a result of the
avalanche. The total number of these dots is then the quantity we
measure as the size of the avalanche.

\section{Results}

\begin{figure}[h]
\centerline{
        \psfig{figure=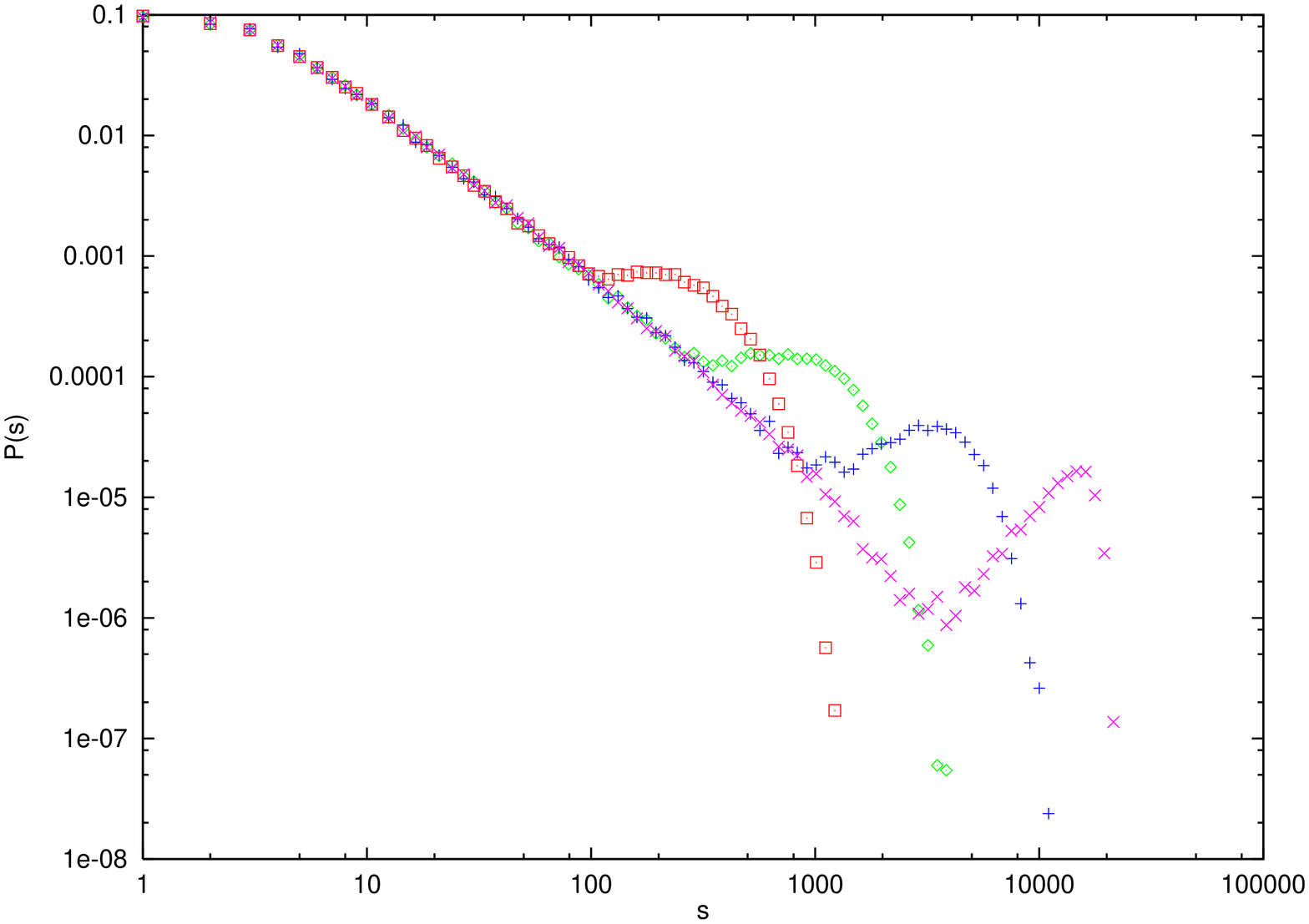,width=12cm,angle=-90}}

        \vspace*{0.5cm}
\caption{}
\label{figloose}
\end{figure}

\begin{figure}[h]
\centerline{
        \psfig{figure=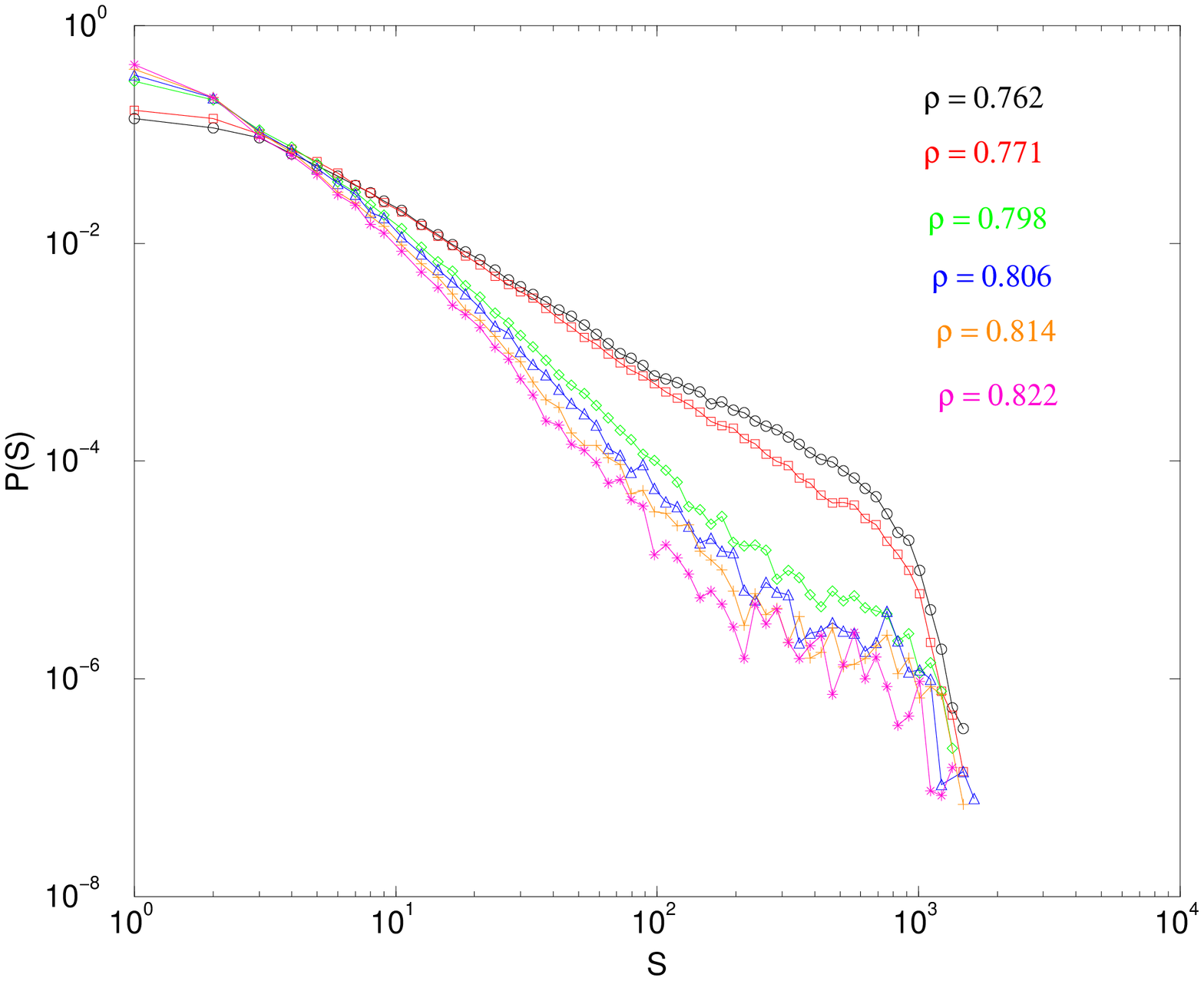,width=12cm,angle=-90}}

        \vspace*{0.5cm}
\caption{}
\label{fighigh01}
\end{figure}

In studying the avalanche distribution as a function of the tapping
density, we used two different values of
$p_{up}$ ( $p_{up}=0.1$ and  $p_{up}=0.5$) in order to test the
sensitivity of the results to different
procedures. 
The avalanche size distribution for the loose-packed density is shown in 
Figure ~\ref{figloose} as a function
of the size of the system. As can be seen, it follows a power-law 
$ P(s) \sim s^{-\tau}$ with an avalanche exponent $\tau$ close to $1.5$. 
The  distribution for the higher densities, according to the tapping procedure 
with $p_{up}=0.1$ and  $p_{up}=0.5$, are shown in 
Figures ~\ref{fighigh01} and ~\ref{fighigh05} for a single system size
($L_x=200,L_y=200$).

The density dependence of the avalanches is highly non-trivial. Since
the loose packed density is the lowest that the packing can reach,
shaking, as mentioned earlier, results in a monotonic increase of the
density. It seems evident from Figure ~\ref{figloose} and Figures
~\ref{fighigh01} and  ~\ref{fighigh05} that the
loose-packed density and other densities close to this value seem to
exhibit a power-law behavior for the avalanche size distribution. 
At very high densities, when the structure is 
ordered (due to the ground state being a completely ordered one)
the avalanche 
distribution is exponentially distributed (as will be clear later).  
Though this
information is insufficient to infer the behaviour of
intermediate densities, a possible hypothesis is that
there exists a second order  
critical point located at some density $\rho_c$. Then
one would expect that for densities larger than the critical one 
the system develops a characteristic size for the avalanches that acts 
as a cut-off for the avalanche size distribution. From this point of view 
one would then expect to be able to rescale all the avalanche size 
distributions obtained at different densities in one single scaling 
function such as
\begin{equation} 
P(s,\rho)= s^{-\tau}F(s{(\rho-\rho_c)}^{1/\sigma})
\label{eq:scaling}
\end{equation}
where $\rho_c$ represents the location of the critical point and
$P(s,\rho)$ is the probability for avalanches propagating in a medium of 
density $\rho$.

\begin{figure}[h]
\centerline{
        \psfig{figure=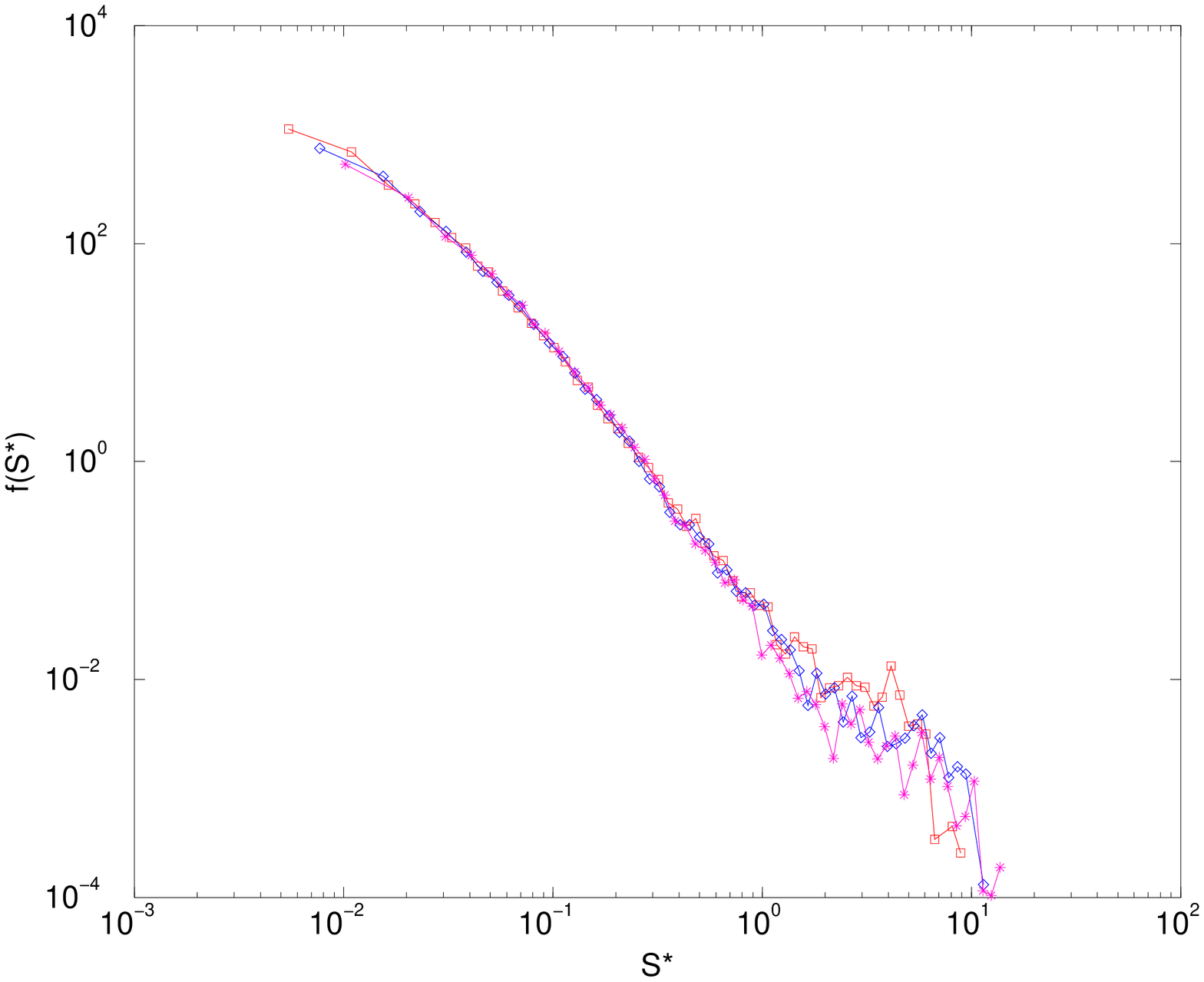,width=12cm,angle=-90}}

        \vspace*{0.5cm}
\caption{}
\label{figcoll_tap01}
\end{figure}

\begin{figure}[h]
\centerline{
        \psfig{figure=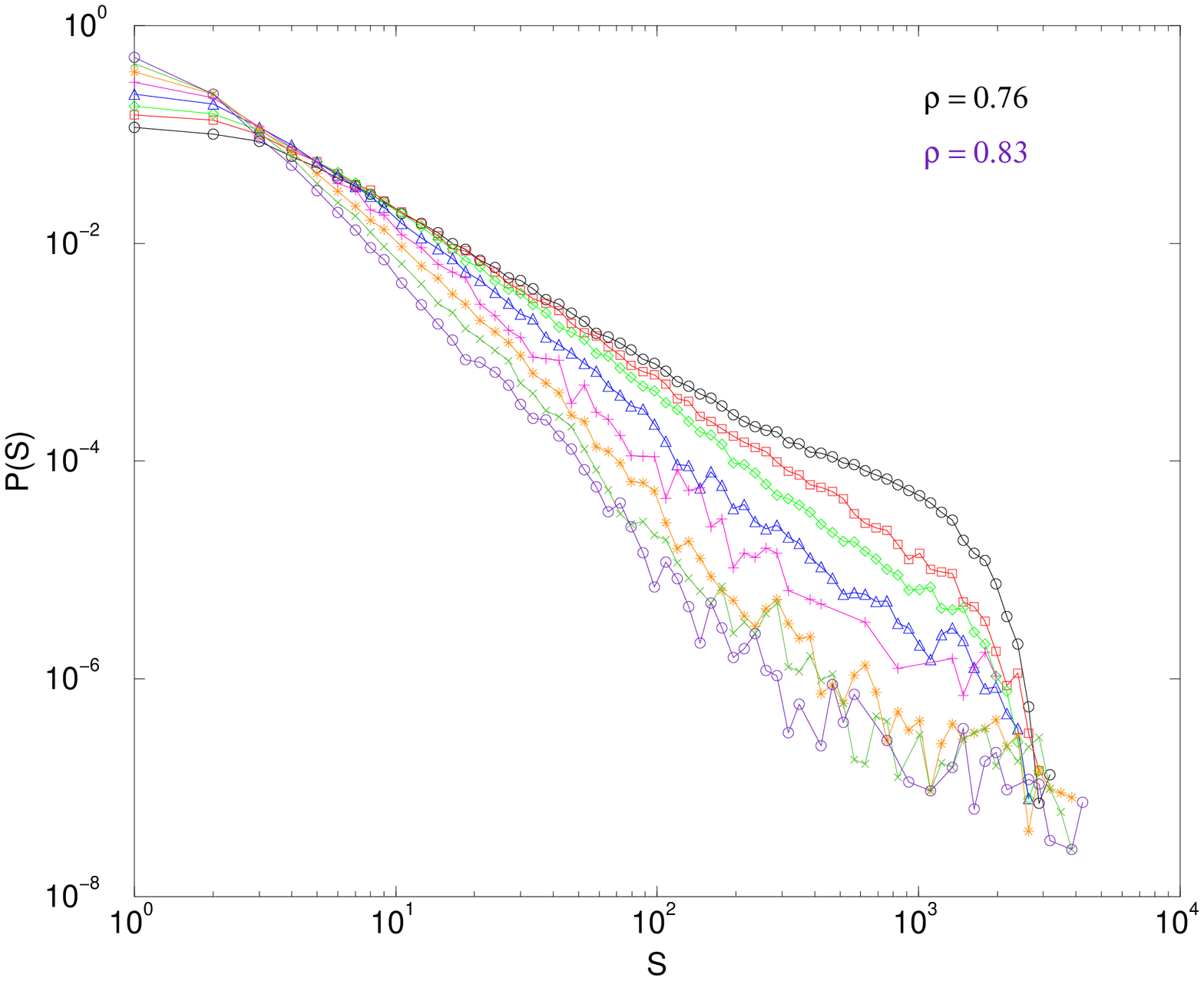,width=12cm,angle=-90}}

        \vspace*{0.5cm}
\caption{}
\label{fighigh05}
\end{figure}

\begin{figure}[h]
\centerline{
        \psfig{figure=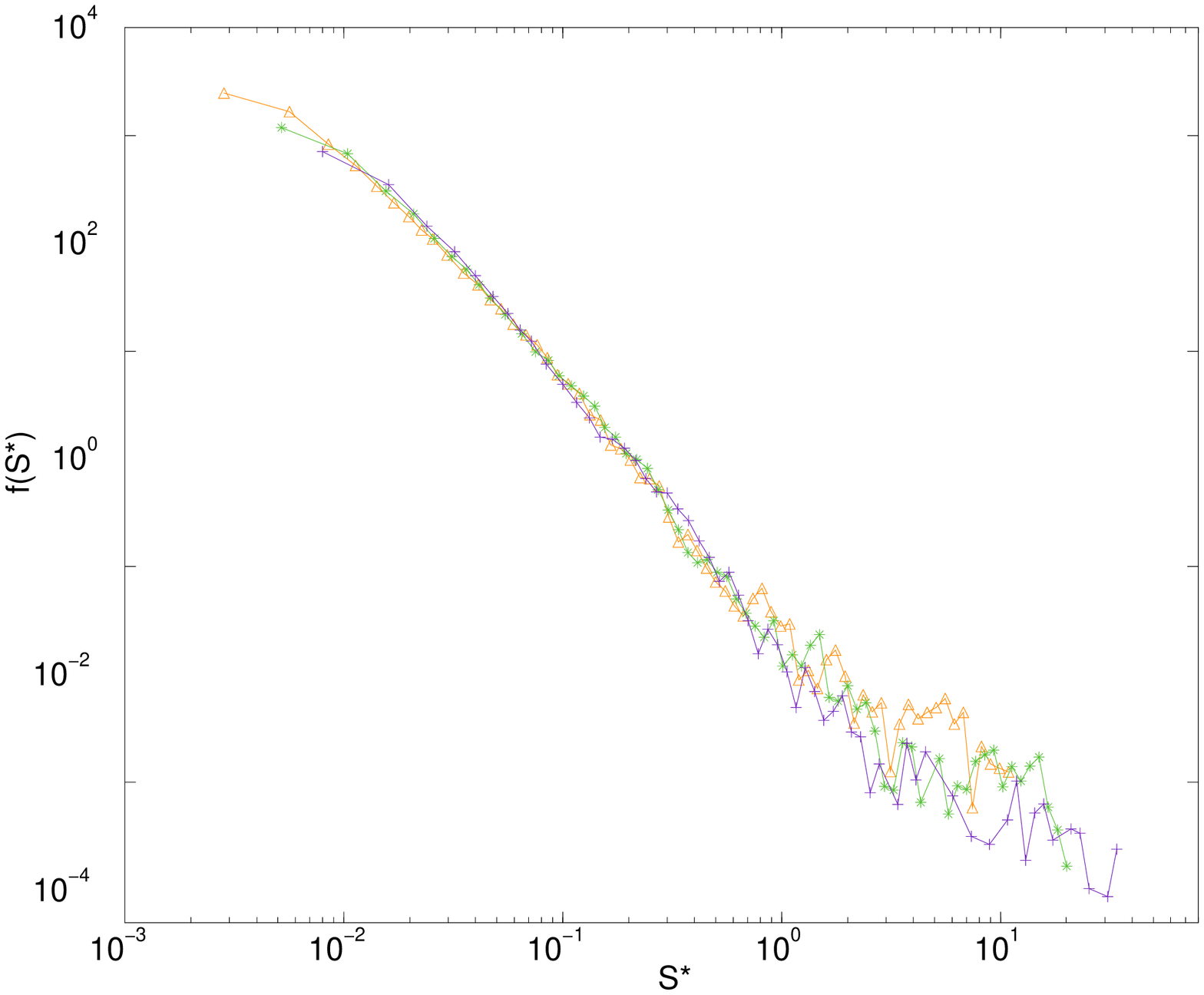,width=12cm,angle=-90}}

        \vspace*{0.5cm}
\caption{}
\label{figcoll_tap05}
\end{figure}

If we make this hypothesis of a single critical
density even for the data, then the 
results for scaling the avalanche data in 
Figures ~(\ref{fighigh01}) and (\ref{fighigh05}) 
are shown in Figures  ~\ref{figcoll_tap01} and
~\ref{figcoll_tap05} respectively. In each case, it is only the last
three curves of the avalanche data that are scaled since the value of
the critical density lies in between the lowest and the highest
we have measured.
As indicated the best value of the exponents
$\tau$ and $\sigma$ seem to match for the two sets of data 
within the error bars. As for the values 
of $\rho_c$, though they seem to depend on the particular procedure used to 
generate the avalanches, we cannot rule out the possibility that they 
coincide within the error bar of our numerics. 

In order to 
test this hypothesis of a single critical density 
in a simpler situation as well as elucidate the possible
meaning of a critical density,
we have also
looked at a toy model which is a simple limiting case of the 
more general situation.  
As already mentioned, the {\em Tetris} model with rods of two
orientations has a very simple ground state (highest density state) - 
the completely antiferromagnetic one. 
We take advantage of this fact by constructing the toy model  in
the following way. We begin with a  $\rho =1$  completely
antiferromagnetic state and generate lower density states by randomly 
removing particles. After each removal, the system is allowed to 
re-settle into a stable state via the avalanche dynamics already 
described. As a further simplification, we consider periodic boundary 
conditions in both $X$ and $Y$ directions.  
This  allows us to eliminate system size effects as well as edge effects 
on the avalanche statistics.
We call this example the Fully Periodic Single Domain (FPSD) model. 

\begin{figure}[h]
\centerline{
        \psfig{figure=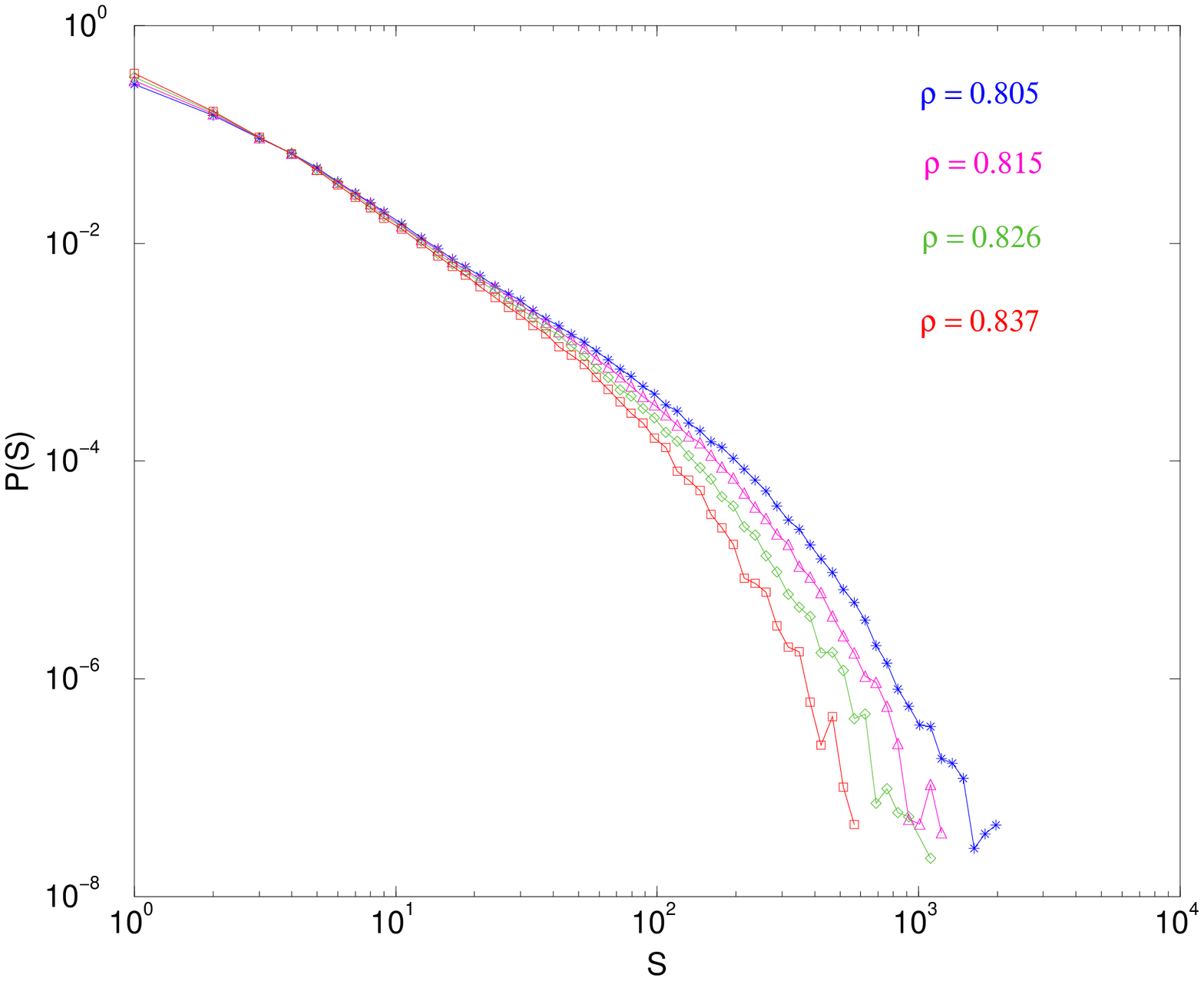,width=12cm,angle=-90}}

        \vspace*{0.5cm}
\caption{}
\label{figsingdom}
\end{figure}

\begin{figure}[h]
\centerline{
        \psfig{figure=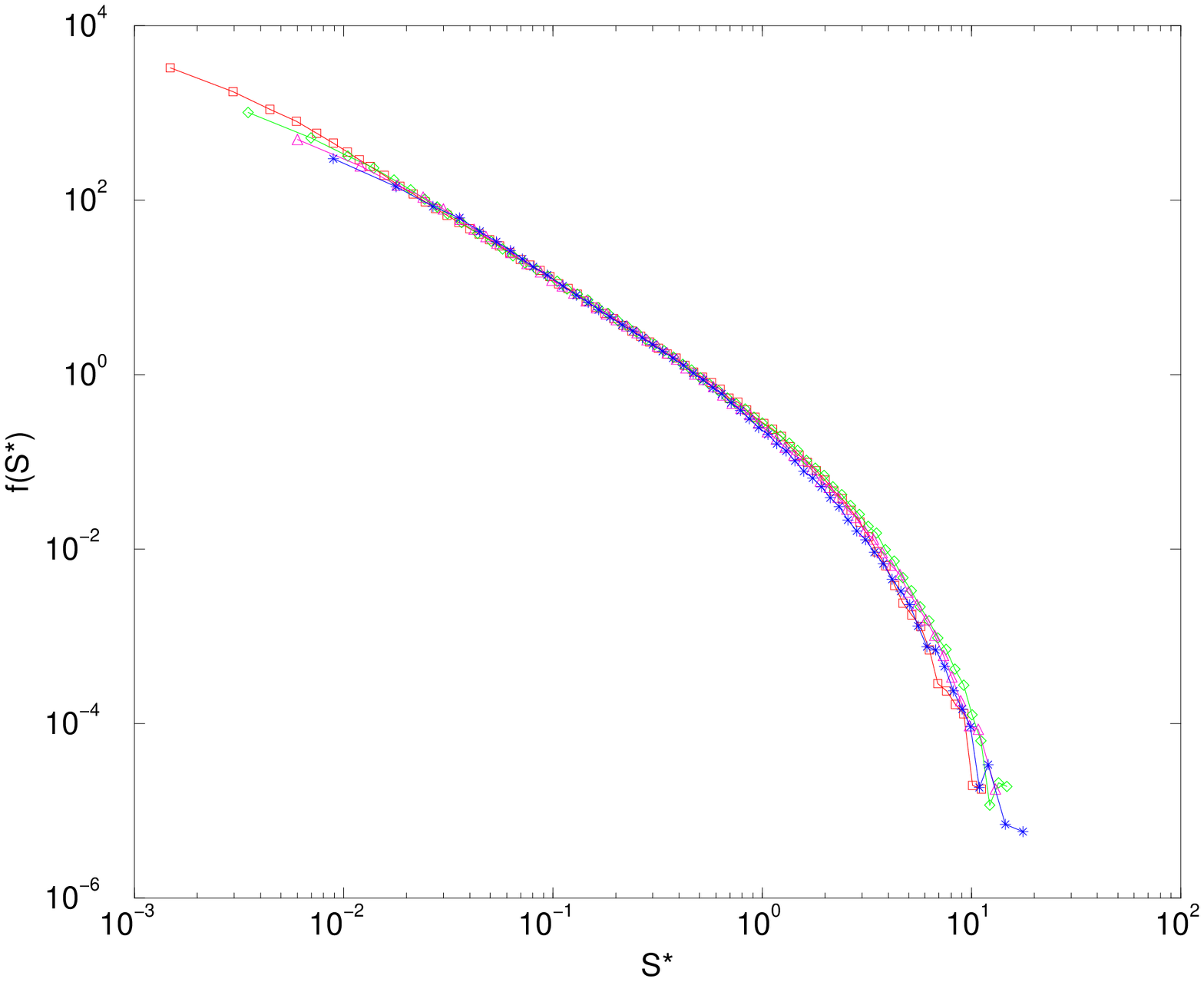,width=12cm,angle=-90}}

        \vspace*{0.5cm}
\caption{}
\label{figcoll_singdom}
\end{figure}

On this simplified version of the model, we perform the same set of
measurements described earlier, in order to measure the avalanche
distribution as a function of the density. A given density here is
accessed by the removal of a certain number of particles instead of by
shaking. Once a given density is reached, the avalanche distribution
is measured by randomly removing a particle from the system and
counting the number of particles destabilized as a result. A
distribution is obtained by the same procedure of averaging over a
ensemble of systems each originating from a $\rho =1$ state from which a
given number of particles (corresponding to the density we want to access)
is removed. The numerical results for avalanche size distribution as a
function of 
the density is shown in Fig. ~\ref{figsingdom}.
In this case the scaling hypothesis is satisfied and
all the curves collapse for a critical density 
$\rho_c^{SD} \sim 0.76 \pm  0.01$, with an avalanche distribution decaying 
as a power with the exponent 
$\tau^{\prime} = 1.45 \pm .05$ and $1/\sigma = 1.5 \pm 0.1$ 
(see Fig.~(\ref{figcoll_singdom})). As can be seen,
the values of the
scaling exponents are 
similar to those obtained for the
original data.

The rationale for the
existence of a
critical point can be understood by
mapping 
the avalanche to a problem of a branching process. 
The avalanche seed represents the insertion of a vacancy in
the system and the subsequent evolution of the avalanche is just the
propagation of this vacancy. There is
an effective probability $p_1 (\rho)$ for the vacancy to 
move through the medium, a probability $p_0 (\rho)$ for stopping
and a probability $p_2 (\rho)$ for branching, {\it i.e.}
meeting some other frozen vacancy and freeing it.
In this mean field description, the
critical point lies at the density which satisfies the
condition $p_2(\rho)= p_0(\rho)$. 
In the FPSD case considered this mapping can be made quantitative by
identifying the configurations leading to branching and death. 
It is possible to have in this way a mean-field estimate of the critical 
density, whose value is in good agreement with the numerical data
presented in Fig.~(\ref{figcoll_singdom}). 

The FPSD does not fully reflect the complexity of the original
problem since we basically probe an ordered state here while the 
configurations ensuing from shaking the loose-packed density (used to
generate  
Figures ~\ref{fighigh01} and ~\ref{fighigh05}) are
disordered. 
The value of the critical density in the FPSD signifies the ordered
loosest 
packing that a single domain state with  periodic boundary conditions
can achieve. The significance of a critical density in the ``shaking''
case is not so clear. The data seems to suggest a density larger than 
the loose packed value. However as mentioned earlier,
we are unable to conclusively pick out a critical density in the latter 
case due to the large error bars. 

It is obvious from Fig.~(\ref{figcoll_singdom}) that the quality of
the collapse is much better than in the case with the open boundaries. 
The error bars on the
various parameters of the collapse are also reduced. This seems to
imply that boundary conditions and surface effects play a 
dominant role in affecting avalanches. One reason that they might do
this is the following. In the FPSD case, by virtue of the boundary
conditions, the system is homogeneous and more or less has the same
density everywhere. Whereas in the curves generated by shaking, the
system develops a
density profile with an interfacial region towards the top where the
density decays to zero. This affects avalanche statistics by enhancing
avalanches over a certain size. Avalanches which reach the low
density region towards the top can very quickly move right to
the surface. The width of the density profile can be reduced by
tapping very gently, {\it i.e.} shaking with a very small amplitude. 
This is possibly an important point to be considered 
and taken into account when doing real life experiments in the
laboratory to study internal avalanching.

\section{Conclusions and Discussion}

Our main conclusions are the following. The avalanche distribution
is very 
sensitive to the internal structure of the packing and clearly 
appears to be long-ranged and decaying as a power law for a range 
of densities.
As obvious in Figs. (~\ref{fighigh01}) and (~\ref{fighigh05})
a slight change in the density changes the avalanche
distribution considerably. However as evident from the figures, 
the avalanche size cut-off is dominated by the system size over a
range of densities, indicating that they are all effectively critical
for the system sizes measured.
Scaling behavior is thus visible only for the
largest densities we have studied. The cleanest case, the case of
the FPSD toy model seems to indicate that there is a critical density
about which other densities obey the scaling relation
Eq. ~\ref{eq:scaling} (Figs. ~\ref{figsingdom} and
~\ref{figcoll_singdom}). 
This might indicate one of two possibilities. The first is 
that there is a unique critical density $ \rho_c \geq \rho_{ld} $ 
even in the case of Figs. ~\ref{fighigh01} and ~\ref{fighigh05}
and we have to go to  systems large enough that the density
dependent cut off appears clearly in the avalanche distribution.
A second possibility is that the avalanche distribution and hence the
critical density crucially depends on the initial state begun with as
well as the procedure under which the system reaches higher densities
(as for example the case considered here, the case of different values
of $ p_{up}$). Our numerical data thus far is unable to distinguish
conclusively between these two possibilities. 

In order to understand the role that the dynamics under which the
system evolves plays in determining the internal structure of the
medium (and hence the avalanche distribution)
we have also studied the following dynamics 
as an alternative to
shaking the system. We begin as usual with the loose packed density
and take out a particle from the bottom row, hence initiating an
avalanche. After the avalanche terminates, we add  the particle back
to the system at a random position on the top.
This dual procedure of removing a particle
and adding it back randomly to the top is continued
till the system reaches a steady state in which the avalanche
distribution is studied. We find that in this case, the steady state
the system reaches is different from any reached in the packing
procedure in that the avalanche distribution decays with a different
power $ \tau =1.75 \pm 0.05$. Further details of this procedure as
well as a description of the steady state reached 
are reported elsewhere \cite{sketal}. 

There are several possibilities which remain to be
investigated. Amongst the most important of these is how much of this
scenario holds for similar models with particles of more complicated
shapes. An example is the {\it Tetris}-like model with ``T''-shaped
particles \cite{tetris,segtet}.  
As mentioned earlier, this model has a highly degenerate ground state 
unlike the simpler model we have studied in this paper.
It is of interest to investigate whether this property
affects the scenario discussed above for the avalanche distribution as
a function of density. While preliminary results indicate that the
avalanche distribution is again a power-law in this case, we have yet
to investigate in detail its density or dynamics dependence. 

\section*{Acknowledgements}
We would like to thank S. Zapperi for useful comments.
We are grateful to CEFIPRA for their support. In
particular SK would like to acknowledge financial support under
project no 1508-3/192. VL acknowledges financial support under project
ERBFMBICT961220. This work has been partially supported
from the European Network-Fractals under 
contract No.FMRXCT980183.

\newpage

\centerline{Figure Captions}

\noindent
{\bf Fig. 1} \, An example of a stable configuration on which
internal avalanche measurements can be made. If the circled particle
is  removed, from the lower most layer, it at most destabilizes all the 
particles within the cone shown. The boundary conditions are
periodic in the horizontal direction ($X$). \\ \\

\noindent
{\bf Fig. 2} \, A picture of an avalanche. The shaded dots represent
all the 
particles in the initial configuration which moved as a result of
the removal of a particle from the lower most layer. \\ \\

\noindent
{\bf Fig. 3} \, Avalanche size probability distribution $P(s)$ 
for the loose-packed density for different system sizes. In
increasing order, the system sizes are $Lx=100,Ly=100$,
$Lx=200,Ly=200$, $Lx=200,Ly=400$ and $Lx=100,Ly=900$. Average over
$100,000$ realizations. \\ \\

\noindent
{\bf Fig. 4} \,
$P(s)$ vs. $s$ for the densities mentioned for
$p_{up}=0.1$. \\ \\

\noindent
{\bf Fig. 5} \,
The scaling plot $f(s^{*})$ vs. $s^{*}$
of the data shown in Fig. 4 where $s^{*} = s (\rho -
\rho_c)^{1/{\sigma}}$ and $ f(s^{*}) = (s^{*})^{-\tau} F(s^{*})$.
The last 
three densities are scaled with parameters $\tau= 1.5 \pm 0.1$,
 $1/{\sigma} = 1.5 \pm 0.1$ and $\rho_c=0.77 \pm 0.1$ \\ \\

\noindent
{\bf Fig. 6} \,
$P(s)$ vs. $s$ for the densities $\rho =0.76-0.83$ for
$p_{up}=0.5$.  \\ \\

\noindent
{\bf Fig. 7} \,
The scaling plot $f(s^{*})$ vs. $s^{*}$
of the data shown in Fig. 6 where $s^{*} = s (\rho -
\rho_c)^{1/{\sigma}}$ and $ f(s^{*}) = (s^{*})^{-\tau} F(s^{*})$.
The densities scaled are $\rho =0.81-0.83$ of the
data shown in Figure 6. The scaling parameters are $\tau= 1.5 \pm 0.1$,
 $1/{\sigma} = 1.5 \pm 0.1$ and $\rho_c=0.79 \pm 0.1$ \\ \\

\noindent
{\bf Fig. 8} \,
$P(s)$ vs. $s$ for the densities mentioned for
the FPSD model. \\ \\

\noindent
{\bf Fig. 9} \,
The scaling plot for the data in Fig. 8. with the scaling
parameters $ \tau =1.45 \pm 0.05$, $1/{\sigma} = 1.5 \pm 0.1$ and $\rho_c
= 0.76 \pm 0.01$ \\ \\

\newpage

\end{document}